# Clarifying the conceptual dimensions of representation in neuroscience[#]

Stephan Pohl [1,*], Edgar Y. Walker [2], David L. Barack [3], Jennifer Lee [4], Rachel N. Denison [5], Ned Block [1], Florent Meyniel [6,7,9], Wei Ji Ma [4,8,9]

[1] Department of Philosophy, New York University, New York, NY, USA.

[2] Department of Neurobiology and Biophysics, Computational Neuroscience Center, University of Washington, Seattle, WA, USA.

[3] Departments of Neuroscience and Philosophy, University of Pennsylvania, Philadelphia, PA, USA.

[4] Center for Neural Science, New York University, New York, NY, USA.

[5] Department of Psychological & Brain Sciences, Boston University, Boston, MA, USA.

[6] Cognitive Neuroimaging Unit, INSERM, CEA, CNRS, Université Paris-Saclay, NeuroSpin center, Gif/Yvette, France.

[7] Institute for Neuromodulation, GHU Paris psychiatrie et neuroscience, Sainte Anne Hospital, Paris, France.

[8] Department of Psychology, New York University, New York, NY, USA.

[9] FM and WM jointly supervised the project.

[*] e-mail: stephan.pohl@nyu.edu

## Abstract

Despite the centrality of the notion of representation in neuroscience, the field lacks a unified framework for the concepts used to characterize representation, leading to disparate use of both terminology and measures associated with it. To offer clarification, we propose a core set of conceptual dimensions that characterize representations in neuroscience. These dimensions describe relations between a neural response, features that may be represented, and downstream effects of the neural response. A neural response may be shown to be *sensitive* or *specific* to a feature, *invariant* to other features, or *functional* (it is used downstream in the brain). We use information-theoretic measures to illustrate these conceptual dimensions and explain how they relate to data analysis methods such as correlational analyses, decoding and encoding models, representational similarity analysis, and tests of statistical dependence or adaptation. We consider several canonical examples, including models of the representation of orientation, numerosity, and spatial location, which illustrate how the evidence put forth in support or criticism of these models is systematized by our framework. By offering a unified conceptual framework to characterize representation in neuroscience, we hope to aid the comparison and integration of results across studies and research groups and to help determine when evidence for a neural representation is strong.

# 1. Introduction

Humans and other animals perceive features of their environment, integrate sensory information with knowledge about rewards and goals, and produce behavior to achieve goals in their environment. Neuroscience aims to explain how the brain brings about these abilities. In a widely used explanatory framework, sensory inputs are transformed into representations of features, while computations transform these representations and integrate them into cognitive processes including reasoning, planning, learning, and memory (Marr, 1982). These processes result in a course of action, which is transformed into motor commands, and executed as the individual’s behavior (Barack & Krakauer, 2021; Perkel & Bullock, 1968).

Neuroscientists agree that the notion of representation is a vital component in their explanatory practices (Baker et al., 2022; Favela & Machery, 2023). Appeals to representations appear frequently. Marr (1982) discusses the transformation of representations of local retinotopic properties to representations of shape properties used for object recognition in vision. Salzman, Britten, and Newsome (1990) find that stimulation with microelectrodes enhances the sensory representations of motion direction. Haxby et al. (2001) suggest that the representations of faces and objects are distributed and overlapping. Kriegeskorte and Diedrichsen (2019) discuss models of the representational geometry of neural responses. Yet, despite this widespread use of representational terminology, neuroscientists report high uncertainty about what kinds of findings count as evidence for representation (Favela & Machery, 2023). The diverse set of tools used to characterize representations and the terminology to describe them leave room for clarification.

In this Perspective, we develop a framework which systematizes the kinds of findings neuroscientists point to in support or criticism of claims of the form that some neural response represents a feature of the environment. Findings can be categorized according to which kind of relation they established between a neural response, features which may be represented, and downstream computations integrating that neural response.

We identify four kinds of relations, which we call the conceptual dimensions of representation. These dimensions are tools for understanding and describing how different methodological approaches contribute to a common project of establishing claims about representations. Experimental designs and data analysis methods can be categorized by the dimensions they provide evidence for. By distinguishing four relations relevant to representation, we describe a space in which findings can be located. This is not to say that any region in this conceptual space corresponds to stronger evidence than other regions, but it is helpful to be explicit about whether two bodies of work address the same aspects of representation. For ease of exposition, we are focused on representations that carry information about the state of the world, whether past, present, or future. Not all representations do. Imaginings, hypothetical reasoning, and representations of goals can occur independently of what the state of the world is.

The conceptual dimensions of representation that we propose are *sensitivity*, *specificity*, *invariance*, and *functionality*. Take the claim that the fusiform face area (FFA) represents faces (Kanwisher et al., 1997). Findings in support of a claim that a given neural response (FFA neural activity) represents a given feature (faces) may show that, first, the neural response is *sensitive* to the feature, that is, it carries information about the feature. Second, they may show the neural response is *specific* to the feature, that is, most variations of the neural response occur only when the feature is changed. Third, findings may reveal that the neural response to the feature of interest is *invariant* to other features; a neural response representing a face, for instance, responds to faces irrespective of its low-level visual features such as position or lighting. Last, they may show the neural response is *functional*, that is, it makes information about the feature available for integration with other cognitive processes, for instance the transformation into motor commands in the form of a behavioral response. The representation of faces in FFA is a classic example of representation in neuroscience. Early papers that developed methodology to identify populations of neurons based on their role in information processing (Kanwisher et al., 1997) and to analyze distributed representations (Haxby et al., 2001) studied the FFA. Sensitivity, specificity, invariance, and functionality have all been addressed in the literature on representations of faces in FFA.

We introduce an information-theoretic formalization of these conceptual dimensions of representation not as a new approach but rather as a conceptual tool to make explicit dimensions systematizing what neuroscientists have been doing all along in studying representations. We develop a framework that researchers can use to integrate results across studies and data analysis techniques. The information-theoretic formalism we use builds on a long line of previous work. Although originally developed in the context of engineering artificial information processing systems (Shannon, 1948), information theory has also been used to analyze natural information processing systems in neuroscience (Perkel & Bullock, 1968) and philosophy (Dretske, 1981; Usher, 2001). It has been used in neuroscience to characterize the information captured by decoding models (Quiroga & Panzeri, 2009), the efficiency with which spike trains represent features (Bialek et al., 1991, 1993; Laughlin, 2001; Rieke et al., 1993), the information transferred within the brain (Ince et al., 2015; Panzeri et al., 2017; Pica et al., 2017; Varley et al., 2023), and it has been proposed as a general statistical framework for neural data analysis (Borst & Theunissen, 1999; Ince et al., 2017; Ostwald & Bagshaw, 2011; Panzeri et al., 2008). Here, we do not endorse particular statistical techniques but instead offer conceptual clarification of research questions about representation to unify common operationalizations in terms of decoding or reconstruction models (deCharms & Zador, 2000; Zhang et al., 1998), brain signatures (Kragel et al., 2018), representational similarity (Haxby et al., 2014; Kriegeskorte & Diedrichsen, 2019) or information theoretic data analyses (Panzeri et al., 2017; Rieke et al., 1999) that are used to address these research questions.

In this Perspective, we begin by introducing sensitivity, specificity, invariance, and functionality in information-theoretic and causal terms. We then consider how common methodologies are used to evaluate the conceptual dimensions of representation, thereby explaining the integration of these methodologies into one unified project. We next illustrate how tests of the dimensions combine to form comprehensive bodies of evidence by looking at several canonical examples of representations: orientation in V1, numerosity in the parietal cortex, and spatial location in the hippocampus. Finally, we briefly consider limitations and potential extensions of our framework of the conceptual dimensions of representation.

# 2. The conceptual dimensions of representation

We introduce our framework using a toy example: we are interested in how the ripeness of an apple is represented in the brain. In our experimental setup, apples are presented to a participant, brain activity is recorded, and the participant chooses whether to eat the apple (Fig. 1). All the ways in which our experimental environment could be different — including the stimulus presented to the participant (apples) — can be summarized in a feature space in which each dimension of that space is one feature. The feature of interest (*s*) — in our experiment, the ripeness of an apple — is distinguished from any other feature (*n*) by being the investigation's target feature. We are testing the hypothesis that the neural response (*r*) represents *s*. The behavioral response (*b*) is the participant's decision to either eat or not to eat the apple (Fig. 1).

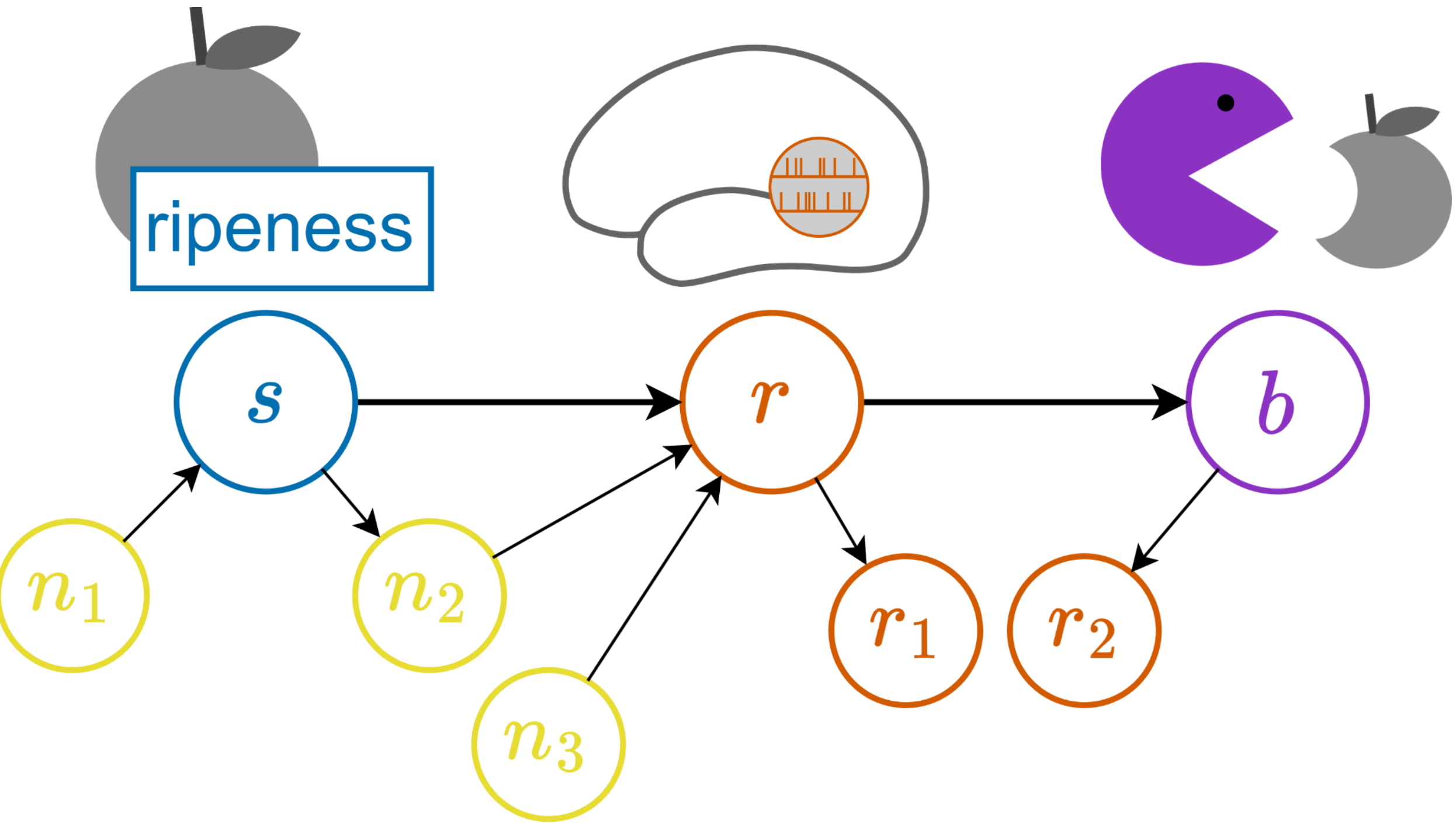

**Figure 1. Generative model of our toy example.** *s* is the ripeness of an apple, *r* is a neural response that might be a representation of *s*, and *b* is a behavioral response; here our participant eats or does not eat the apple depending on whether they perceive it to be ripe. To show that *r* is a representation of *s*, alternative hypotheses about other features (*n*) are considered that stand in diverse statistical relations to *s* and *r*. In our toy example, *n* is exemplified by the age of an apple ($n_1$), the color of an apple ($n_2$), or the color of the packaging in which the apple is presented to the participant ($n_3$). In addition, *r* also is shown to be used as a representation of s. A brain area into which activity from *r* spills over ($r_1$), but which has no role in *s* related processing, or areas involved in monitoring the participant's own behavior ($r_2$), may be related to *s* but fail to be functional

as a representation of *s*. In the figure, nodes represent variables and arrows direct causal dependencies which determine statistical dependencies between variables.

In typical experimental contexts, *s* could be any feature of interest, such as the orientation of a grating, the category of an object, or a more abstract feature such as the probability with which another feature changes its value. Meanwhile, *n* could be any other feature included in the analysis, such as the contrast of an image, the perspective from which an object is viewed, or the modality of a stimulus or the task context. Often, multiple features are included, so that *s* and *n* are multi-dimensional. In our toy example, we identify three features $n_1$, $n_2$, and $n_3$, which differ in how they are causally and statistically related to *s* and *r* (Fig. 1). Going forward, we collapse these into one multi-dimensional variable *n*. Typically, *r* is a high-dimensional vector reflecting the activity pattern of a population of neurons, for example measured with functional MRI (fMRI), electroencephalography (EEG), or implanted electrodes. Finally, *b* could be any behavioral response to *s*, such as an eye movement or a button press by which the participant, for instance, reports the value of *s* or makes some decision based on *s*.

Some methodological approaches are not framed in terms of the representation of features, but in terms of representations that are selective for a target category. A neural event may indicate the presence of a member of that category, such as the firing of neurons in the FFA indicating the presence of faces (Kanwisher et al., 1997). Note that a population of neurons that is selective to a target category, thereby is also sensitive and specific to the binary categorical feature that can take as its value either that the target category is instantiated or not. Modeling the representation as a representation of features can also be extended to many-valued categorical features, such as object category, or to features with a graded scale, like orientation. For many-valued categorical features, it does not make sense to speak of the presence or absence of a target category, which is why we prefer the more general framing in terms of features (Fig. 1).

Often, experiments start with *s* and search for that *r* in terms of which *s* is represented (for instance, to investigate where in the brain faces are represented (Kanwisher et al., 1997)). By contrast, some start with a given *r* and try to find which *s* is represented (for instance, to investigate the receptive field of V1 neurons (Hubel & Wiesel, 1959), or by using model-based approaches that fit models of *s* to *r* (Dumoulin & Wandell, 2008). Both approaches can be combined such that models of neural sub-populations (*r*) and feature spaces (*s*, *n*) are advanced in unison (Chang & Tsao, 2017).

The evaluation of the conceptual dimensions of representation — sensitivity, specificity, invariance, and functionality — depends on which features are included in the analysis and so it may be misleading when relevant features are excluded. In our toy example, we are looking for a representation of ripeness but say we do not consider color as an additional feature and test our hypothesis with a species of apple where red apples tend to be ripe but green apples are not. By excluding a relevant feature (color) from our analysis, we may identify some *r* that is sensitive to ripeness (*s*) even though it is a representation of color ($n_2$; Fig. 1). A better experimental setup would modulate ripeness and color independently, by using a species of apple which may be green even when ripe. With color included as a relevant feature, *r* no longer appears to be sensitive to ripeness (when holding fixed the color, *r* no longer carries information about ripeness) and fails to be invariant to color (even while holding ripeness fixed, *r* continues to carry information about color). In summary, whether measures of sensitivity and invariance reveal *r* to be a representation of color or mistake it for a representation of ripeness depends on whether all relevant features are included in the analysis.

We first introduce the conceptual dimensions of sensitivity, specificity, and invariance in information theoretic terms (Fig. 2; see Box 1 for a summary of basic notions of information theory). These dimensions describe the relation between *s*, *n*, and *r*. Functionality describes the causal role of *r* in downstream cognitive processes and the production of *b* and is introduced subsequently.

The order in which we introduce these conceptual dimensions of representation is chosen for didactic reasons. In practice, there is no general order in which the dimensions are evaluated, with findings about one dimension often inspiring tests of others in a complex back and forth.

## 2.1. Relations between features and the neural response

The relation between features and the neural response is described in terms of *r*'s sensitivity to *s*, its specificity to *s*, and its invariance to *n*. Sensitivity and specificity may further be evaluated conditional on *n*.

**Box 1: Information theory**

Here we provide a brief introduction to central concepts of information theory (see Cover and Thomas (2006) for a more extensive treatment).

In the introduction of our framework, we assume that variables are discrete (see section 3 'Our framework applied across methodologies' for discussion of the application of the framework in data analysis, which includes continuous variables). We write random variables as lower-case letters (for example, $x$). Values of the variables are written with an index (for example, $x_i$). $x_i$ also abbreviates the event $x = x_i$.

**Entropy (*H*)**
Entropy measures the uncertainty about a random variable, which can also be understood as the amount of variability or randomness associated with a variable. Entropy is high when a variable has many values with equally high probabilities ($p$).

**Definition:** $H(x) = -\sum_{i} p\left(x_i\right) \log p\left(x_i\right)$

**Conditional entropy**

The conditional entropy measures how much uncertainty there is on average about some variable conditional on another variable — that is, it measures the average uncertainty about the first variable that remains after the value of the second variable is specified.

**Definition:** $H(x|y) = -\sum_i p(y_i) \sum_j p(x_j|y_i) \log p(x_j|y_i)$

**Mutual information (*I*)**

The mutual information between two variables measures how much uncertainty about one variable is reduced by conditioning on the other variable. That is, mutual information is high when there is much less variability of the values of one variable conditional on the other variable than there is variability not conditional on any variable. Mutual information is symmetric ($I(x;y) = I(y;x)$).

**Definition:** $I(x;y) = H(x) - H(x|y) = H(y) - H(y|x)$

Mutual information can also be understood as a measure of the statistical dependence between two variables. When $x$ and $y$ are statistically independent, then $I(x;y) = 0$. When there is some statistical dependence between $x$ and $y$, then $I(x;y) > 0$.

**Joint entropy**

The entropy of a joint distribution is calculated as follows.

**Definition:** $H(x, y) = -\sum_{i,j} p(x_i, y_j) \log p(x_i, y_j)$

**Conditional mutual information**

Conditional mutual information measures the mutual information between two variables conditional on a third variable. That is, it measures how much, on average, uncertainty about a first variable is reduced by conditioning on a second variable, given that the value of the third variable is already specified.

**Definition:** $I(x;y|z) = H(x|z) - H(x|y,z) = H(x,z) + H(y,z) - H(x,y,z) - H(z)$

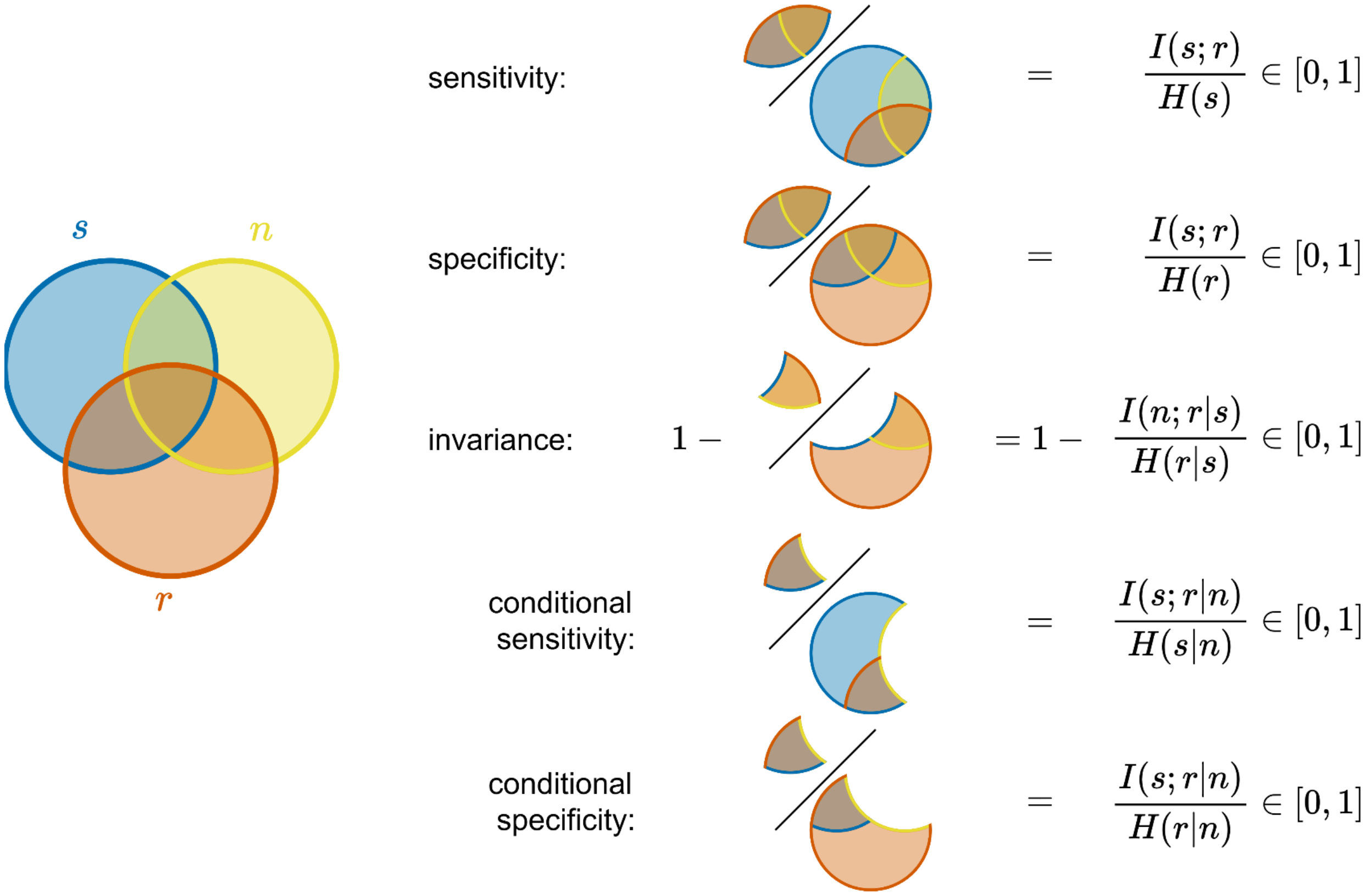

**Figure 2. The notions of sensitivity, specificity, and invariance in information-theoretic terms.** The variables *s*, *n*, and *r* correspond to the feature of interest, other features, and the neural response respectively. Circles in the Venn diagram represent the total variability associated with a given variable measured in terms of the entropy (*H*) of that variable (left). Overlapping areas represent the mutual information (*I*) between variables. Information-theoretic quantifications of sensitivity, specificity, invariance, and the conditional forms of sensitivity and specificity are given by the equations (right). The shapes depict the same quantities in terms of the respective areas in the Venn diagram (middle). The numeric values of sensitivity, specificity, and invariance range from 0 to 1, with 1 indicating maximal satisfaction and 0 indicating no satisfaction.

### 2.1.1. Sensitivity

*r* is sensitive to *s* given that *r* is informative about *s*. In our toy example (Fig. 1), *r* is sensitive to the ripeness of an apple given that ripeness can be inferred from *r* (for example, because the neural response rate is higher for ripe apples than for non-ripe apples). An empirical finding of sensitivity is the observation that neurons in the middle temporal area (MT) respond preferentially to directional motion allowing the feature of motion direction to be predicted from their responses (Dubner & Zeki, 1971).

For $r$ to be sensitive to $s$, there must be mutual information ($I$) between $r$ and $s$ (Box 1). However, the absolute amount of mutual information is not very useful. What matters for sensitivity is that the mutual information between $r$ and $s$ captures a large proportion of the total information about $s$. If $r$ captures the total information about $s$ (such that $I(s;r) = H(s)$)) it is possible to infer the value of $s$ from $r$ without error (Fig. 3a, right). This upper bound on the measure of sensitivity is achieved by normalizing the mutual information between $s$ and $r$ by the entropy ($H$) of $s$ (Box 1; Fig. 2). In statistics, this normalization of mutual information is sometimes called the uncertainty coefficient (Press et al., 2007).

$$\text{Sensitivity:} = \frac{I(s;r)}{H(s)}$$

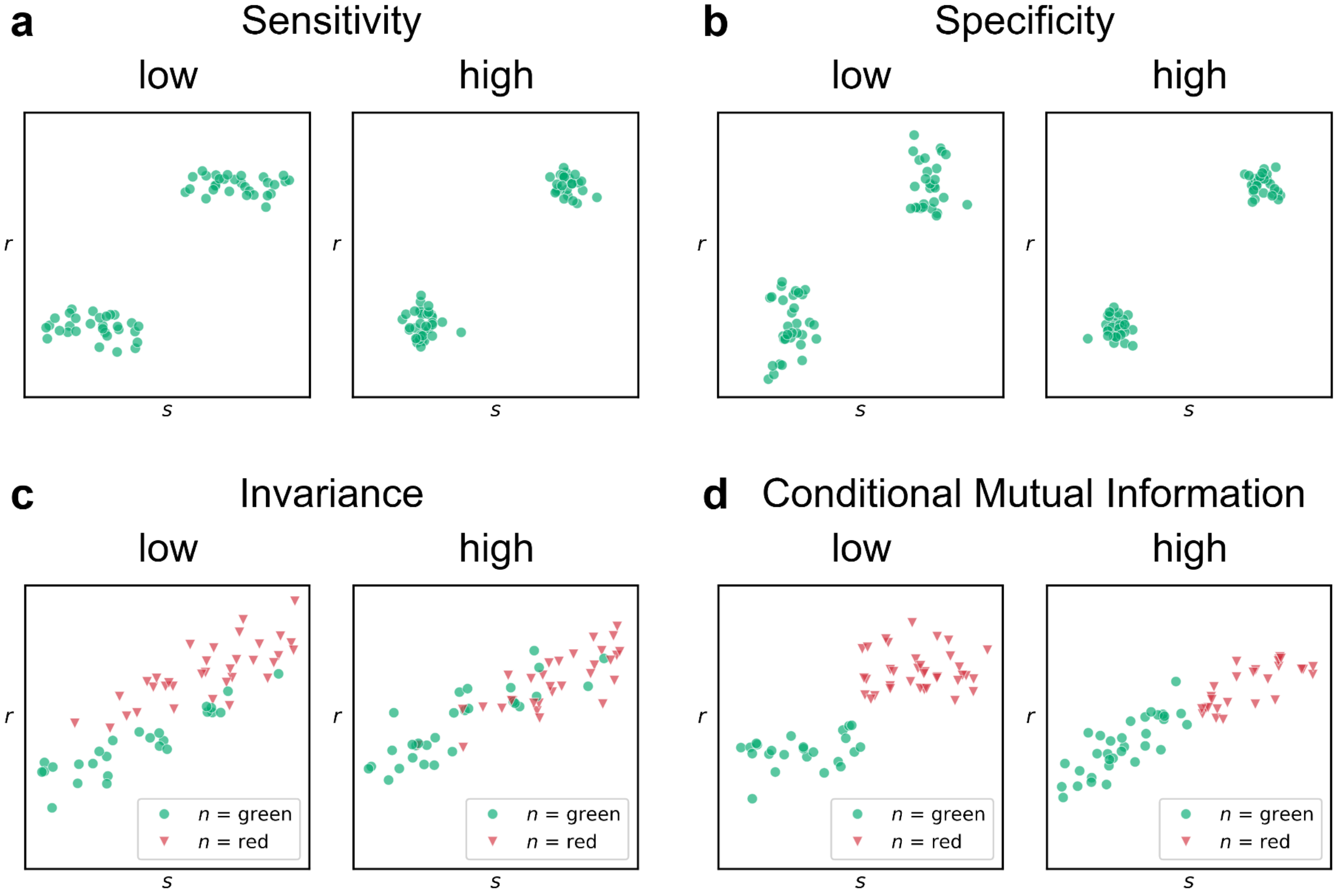

**Figure 3. Sensitivity, specificity, and invariance capture relations between features and the neural response. a** | Sensitivity is high when $s$ can be inferred from $r$ with high accuracy. When sensitivity is low, the distribution of $s$ given $r$ is more spread out such that a decoding model of $s$ given $r$ would be less accurate (left). By contrast, given $r$, little variability of $s$ remains when sensitivity is high, such that a decoding model of $s$ given $r$ can be highly accurate (right). Changes like this in the stimulus distribution across low versus high sensitivity conditions may be less typical for controlled experiments but are to be expected in more

natural environments. **b |** Specificity is high when *s* explains a large proportion of the variability of *r*. When specificity is low, conditional on *s*, the remaining amount of variability of *r* is large (left). Whereas for a given value of *s*, there is little difference between values of *r* when specificity is high (right). In **a**, only sensitivity changes, but specificity remains unchanged, whereas in **b**, specificity changes and sensitivity remains unchanged; in the low conditions, additional variability is introduced which spreads out the distribution either along *s* (sensitivity) or *r* (specificity) but leaves the distribution along the other variable unchanged. Such a clear dissociation of sensitivity and specificity is rare in practice and exaggerated here in our simplified example. Yet sensitivity and specificity do commonly come apart, for instance, when neural noise or other drivers of neural variability undermine specificity whereas sensitivity remains unaffected because the neural code for *s* is robust to the interfering drivers of variability (such as by averaging out neural noise). **c |** Invariance is high when, conditional on *s*, there is no dependence of *r* on *n*. There may be a dependence of *r* on *n*, even though invariance is high, if this dependence disappears conditional on *s* (for a given value of *s*, it is no longer the case that *r* tends to be larger for *n* = red, than for *n* = green; right). When even conditional on *s*, *r* depends on *n* (*r* tends to be larger for *n* = red, than for *n* = green; left), the invariance is low. **d |** The mutual information between *s* and *r* conditional on *n* is high when, even conditional on *n*, there is a strong dependence between *s* and *r*. When conditional on *n*, the dependence between *r* and *s* is broken (for a given value of *n*, *s* and *r* are statistically independent; left), the mutual information between *r* and *s* conditional on *n* is low. When the dependence persists (even conditional on *n*, *r* is informative about *s*; right), the conditional mutual information is high.

That *r* is highly sensitive to *s* therefore means that the information *r* carries about *s* captures a large proportion of the variability of *s*. Because of the normalization, sensitivity takes values between 0 (no sensitivity) and 1 (perfect sensitivity), but perfect sensitivity cannot generally be expected, as representations may be inaccurate. For instance, people inexperienced in telling species of trees apart may represent a tree as being a beech, however, that representation would often be inaccurate (and its sensitivity to the feature of tree species substantially below 1). Despite this, tree species might still be the feature to which that representation is most sensitive.

### 2.1.2. Specificity

*r* is specific to *s* if many of the changes in *r* are explained by changes in *s*. In other words, *s* explains a large proportion of the variability of *r*, or, given *s*, *r* remains mostly constant (Fig. 3b, right). Recall that *s* can be multi-dimensional (in our toy example we might add a second feature of interest besides ripeness, like sweetness; Fig. 1). A representation that is sensitive to multiple features — displaying mixed selectivity (Fusi et al., 2016; Rigotti et al., 2013) — will be specific only to the combination of these features.

In information-theoretic terms, specificity is quantified by the mutual information between *s* and *r*, normalized by the entropy of *r* (Fig. 2).

$$\text{Specificity:} = \frac{I(s;r)}{H(r)}$$

That *r* is highly specific to *s* therefore means that a large proportion of the variability of *r* carries information about *s*. An equivalent expression of specificity is as follows:

$$1 - \frac{H(r|s)}{H(r)}$$

That is, specificity can also be thought of as the proportion of the variability of *r* that is explained by *s*. A lack of specificity suggests that some factors other than *s*, be they noise, internal processes or *n*, drive the remaining variability of *r* (in Fig. 2, $H(r|s)$, the area of $H(r)$ not overlapping with $H(s)$, would be large). Specificity quantifies only the dependence of *r* on *s*, but not the dependence on other factors. A finding of low specificity (Fig. 3b, left) may be followed by tests aiming to identify the remaining factors driving the variability of *r*.

Specificity has also been used under the label 'coding efficiency' (Bialek et al., 1993; Rieke et al., 1993). This work evaluated what proportion of the variability of spike trains carries information about the stimulus; the larger the proportion of the variability of spike trains that carries information about the stimulus, the more efficient is the neural code — the mapping from *r* to estimates of *s* (see Supplementary Box 1). In a neural code of low efficiency, much of the variability of *r* fails to carry information about *s* which may be metabolically costly (Haueis & Colaço, 2025).

Sensitivity and specificity both depend on the mutual information between *r* and *s*; they differ only in the normalizing factor (Fig. 2). High sensitivity means that *s* can be inferred from *r* with high accuracy whereas high specificity means that *r* can be inferred from *s* with high accuracy (Fig. 3a, b)). Often, sensitivity is approximated by the performance of decoding models and specificity is approximated by the performance of encoding models (Naselaris et al., 2011). The close connection between sensitivity and specificity, but also the importance of distinguishing these two relations has been observed previously. Poldrack (2006, 2011) distinguishes between forward inferences (based on encoding models) and reverse inferences (based on decoding models) and points out how the validity of a reverse inference (decoding a representation from *r*) depends on the specificity of *r*. Averbeck et al. (2006) pointed out how noise correlations may have different impacts on how information about features is encoded in *r* (reflecting specificity) or decoded from *r* (reflecting sensitivity).

The notions of sensitivity and specificity in our framework should not be confused with sensitivity and specificity in a binary decision or test context. In those contexts, 'sensitivity' refers to the true positive rate (the proportion of positive instances truly detected as positive) and 'specificity' refers to the true negative rate (the proportion of negative instances truly detected as negative). A high-level analogy exists between the two notions; in both, sensitivity indicates that a test or representation is responsive to its target whereas specificity indicates that it is not responsive to other things. Yet, mathematically, these notions are quite different.

A perfect specificity of 1 is not usually expected to be found because of factors unrelated to *s* that affect the neural response, including neural noise (Faisal et al., 2008) and internal processing reflecting brain states such as arousal, motivation, or mind wandering.

### 2.1.3. Invariance

The representation of *s* by *r* is invariant to *n* if *r* does not depend on *n* for a given value of *s* (if *r* is not sensitive to *n*, conditional on *s*). In our toy example (Fig. 1), to show that the representation of ripeness by *r* is invariant to the apple's color, one would have to show that *r* indicates ripeness independently of color. A typical empirical finding of invariance is that neural responses in the primate inferior temporal cortex, that are sensitive to object category, did not respond to changes in the orientation of the object (Logothetis et al., 1995; Quiroga et al., 2005), or that neural responses, that are sensitive to surface color, remained unchanged across variations in the ambient light color (displaying color constancy) (Zeki, 1980).

When *s* and *n* are statistically related and *r* is sensitive to *s*, it may also be sensitive to *n* in virtue of the statistical dependence between *s* and *n*. So, *r* cannot generally be expected to lack sensitivity to *n* even though it may be invariant to *n*. To demonstrate invariance, one must show that *r*'s dependence on *n* is accounted for by *r*'s dependence on *s*. In other words, *r* is invariant to *n*, if, conditional on *s*, *n* does not account for any of the variability of *r* (Fig. 3c, right).

In information-theoretic terms (Fig. 2), invariance is quantified as follows:

$$\text{Invariance:} = 1 - \frac{I(n;r|s)}{H(r|s)}$$

Which, equivalently, can be expressed as follows:

$$\frac{H(r|n,s)}{H(r|s)}$$

An invariance of 1 means that no more of the variability of *r* can be accounted for by the combination of *s* and *n* than by *s* alone. An invariance below 1 means that some of the variability of *r* that is unrelated to *s* can be explained by *n*.

Two applications of invariance are of particular interest, namely those in which *n* stands for the modality of a stimulus or for the task. A representation of location might be invariant to whether the location of an event is presented to a participant as a visual or an auditory stimulus (*r* is invariant to *n* when *r* represents *s* both in the condition when *n* = auditory and when *n* = visual). Invariance to modality is great evidence for the claim that *r* is a representation of *s*. However, lack of invariance to modality is more difficult to interpret — it might be that the same features are independently represented in the perceptual systems specialized for different modalities.

Another application is task invariance; here *r* is task-invariant when it is sensitive to *s* irrespective of *n*. A lack of task invariance does not need to indicate that *r* is a poor candidate for a representation of *s*; it might be that the brain only represents *s* when it has to. That *r* is particularly sensitive or specific only in conditions where it is task relevant may even suggest that it is used in the response to the task (that it is functional). For instance, in the study of feature-based attention, task context has been found to modulate sensitivity of representations to particular dimensions of the feature space like motion or color (Chawla et al., 1999).

Both the invariance and specificity of *r* for complex features (such as object category (DiCarlo & Cox, 2007) or numerosity (Castaldi et al., 2019)) tend to increase along perceptual processing hierarchies. By the data processing inequality (Cover & Thomas, 2006) — which states that, along a causal chain, mutual information can only be maintained or decreased but not increased — sensitivity can only stay the same or decrease from early to later processing stages. However, neural responses at later processing stages tends to be more specific to complex features and invariant to low-level features (such as retinotopic maps of orientation, contrast, and so on) as representations of the complex features are disentangled from the low-level features (DiCarlo & Cox, 2007). For instance, object category can already be decoded from early visual areas using sophisticated models; however, in inferotemporal cortex populations of neurons are highly specific to object category and invariant to low-level features such as the object's orientation. Relatedly, invariance may often be interpreted as the result of a computation or inference which reconstructs the value of latent features (such as object category) from intermediate features more directly observable by the participant (such as the retinal image) (Burge, 2010; Clark, 2013; Friston, 2010; Helmholtz, 1867; Hohwy, 2013).

### 2.1.4. Conditional sensitivity and specificity

Sensitivity and specificity can be evaluated conditional on *n*. A common goal for testing conditional sensitivity and conditional specificity is to rule out that *r* depends on *s* merely because *r* depends on *n* and *s* happens to also depend on *n*.

In our toy example (Fig. 1), we may test whether even just for green apples, *r* continues to depend on the ripeness of the apple (whether *r* differs between green apples that are ripe and green apples that are not ripe). If *r* were merely a representation of color, conditional on color, it would not carry any information about ripeness (Fig. 3d). A typical empirical test of conditional sensitivity, for instance, may show that a neural response to the numerosity of a dot cloud carries information about numerosity even while holding fixed related quantities like density, shape, or surface area covered by the dot cloud (Nieder et al., 2002).

As with their unconditional analogues (Fig. 2), conditional sensitivity and conditional specificity are quantified by different normalizations of the mutual information between *s* and *r*, but conditional on *n*:

$$\text{Conditional sensitivity:} = \frac{I(s;r|n)}{H(s|n)} \qquad \text{Conditional specificity:} = \frac{I(s;r|n)}{H(r|n)}$$

That *r* is sensitive to *s*, conditional on *n*, means that the information *r* carries about *s*, after conditioning on *n*, captures a large proportion of the variability of *s*. That *r* is specific to *s*, conditional on *n*, means that, after conditioning on *n*, a large proportion of the variability of *r* carries information about *s*. The conditional variants of sensitivity and specificity, as well as invariance, are of particular relevance when *r* depends on interactions of *s* and *n* (Supplementary Box 2).

In practice, sensitivity and specificity are always evaluated conditional on the experimental setup. This may undermine the ecological validity of the findings. For instance, it introduces the risk of finding dependencies conditional on some aspect of the experimental setup that may not generalize (for example because features covary in the experiment but not in the natural world). Or it may obscure dependencies which are present in natural environments but may be absent in constrained experimental setups. Controlling the experimental environment, varying experimental conditions, and randomizing them can mitigate these risks; however, they cannot be avoided completely.

## 2.2. Functionality

If *r* is a representation of *s*, we not only expect it to carry information about *s* but also expect that this information is used in the brain (Baker et al., 2022; deCharms & Zador, 2000; Perkel & Bullock, 1968; Ritchie et al., 2019; Shadlen & Newsome, 1994) (that *r* is functional). Generally, when *r* is a representation of *s*, subsequent processes that rely on information about *s* receive that information from *r*. In our toy example (Fig. 1), *r* is a representation of ripeness when the participant makes the decision to eat an apple (*b*) based on whether *r* represents the apple to be ripe or not (*r* is functional in virtue of being used in this decision).

Use implies causality. Therefore, functionality has to be evaluated in causal terms (Brette, 2019; Jones & Kording, 2019; Panzeri et al., 2017). Causal claims can be established with experimental interventions (Pearl, 2009; Woodward, 2003) — to show that *r* is functional, one can show that interventions on *r* modulate what information about *s* is available to downstream processes (such as a behavioral report of *s*). For example, Salzman et al. (1990) showed that stimulation of neurons with a specific preferred motion direction biases the behavioral report of the perceived motion direction towards that preferred motion direction.

Information-theoretic quantities, much like correlations, are not sufficient to capture causality and, thereby, functionality, but they are still useful. General statistical methods for inferring causal dependencies from observational data have been discussed in the context of cognitive neuroscience (Bressler & Seth, 2011; Friston et al., 2003; Granger, 1969; Valdes-Sosa et al., 2011; Weichwald & Peters, 2021). A precursor to claims about functionality are claims about functional connectivity (claims about statistical relations between populations of neurons that suggest the transfer of activation and information between them) (Bastos & Schoffelen, 2016; Friston, 1994; Luppi et al., 2022; Stephan & Friston, 2009). More specific statistical methods have been developed to trace the transfer of information about a feature between populations of neurons (Celotto et al., 2023; Ince et al., 2015) or to trace the transfer of information from *r* to *b* (Lemke et al., 2024; Purushothaman & Bradley, 2005; Rossi-Pool et al., 2021) (for example, in terms of the probability of correctly predicting *b* from *r* (Britten et al., 1996)). Many methods consider the relations between *s* and *r* or between *b* and *r* in isolation. To address functionality more specifically, it has been proposed to analyze the intersection information (Panzeri et al., 2017; Pica et al., 2017) which quantifies not only the information transfer from *s* to *r*, or from *r* to *b*, but quantifies how much information is in common between *s*, *r* and *b*.

The impact on *b* of most representations is mediated by other representations. For instance, sensory representations may be integrated with representations of goals to form plans for actions and to produce motor commands, which are turned into *b*. A full understanding of functionality requires tracing the flow of information about *s* from sensory input to motor commands and *b* (in the form of multi-stage process models). For example, Wilming et al. (2020) showed in a perceptual decision-making task how sensory signals are represented in early visual areas, accumulated and transformed into action plans in parietal and motor areas, and, ultimately, turned into a behavioral response by pressing a button with the participant's left or right hand. Further, choice reflective motor activity fed back into early visual areas, modulating how much sensory representations are integrated into the decision-making process.

Statistical methods for inferring causal connections are fundamentally limited. A central challenge is that a statistical dependence between two variables need not be due to a causal connection between these variables but may be due to a causal dependence on a third variable. For instance, in our toy example (Fig. 1) there is a statistical dependence between $r_1$ and *b* — they may even share information about *s* — not because $r_1$ causes *b*, but because both share *s* (and, in this case, also *r*) as common causes. So $r_1$ may carry information about both *s* and *b* without being functional. Some tests of functionality try to exclude a shared statistical dependence on *s*, by controlling *s*. For example, Bradley et al. (1998) showed that the reported motion direction of bistable stimuli — 2D projections of rotating transparent cylinders where the perceived motion direction changes spontaneously between the two possible directions of rotation around the cylinder's central axis — could be predicted from neural activity in area MT. Similarly, functionality can be evaluated by showing that errors (variability of *b* that is unrelated to *s*) can be predicted from *r* (Purushothaman & Bradley, 2005).

No observational data can exclude the possibility that some unobserved *r* explains a statistical dependence between variables without a direct causal connection. Where possible, experiments in which an intervention is performed directly on *r* that leave stimulus and task context unchanged are particularly valuable. Changes in *b* following such an intervention are strong evidence of a causal role of *r* in bringing about *b* (evidence of functionality). For example, in non-human primates, stimulation of neurons in area MT modulates the perceived direction of motion, providing evidence that area MT is functional as a representation of motion direction (Salzman et al., 1990), whereas stimulations in area LIP modulate how quickly a participant reports a specific direction of motion, providing evidence that area LIP is functional as a representation of the strength of sensory evidence about motion direction (Hanks et al., 2006). Similarly, chemically induced lesions of area MT in monkeys (Newsome & Pare, 1988) and suppression of area V5 with transcranial magnetic stimulation (tMS) in humans (Beckers & Homberg, 1992) both impaired motion perception specifically. Interventions performed directly on *r* may not address effects on *b* but instead effects on subsequent neural processing, for example, stimulations of neurons in the frontal eye field increased the information about stimulus features carried by retinotopically corresponding neurons in area V4 of non-human primates (Moore & Armstrong, 2003). Most work cited in this paragraph used invasive intervention methods and was performed with monkeys. Only the study using the non-invasive tMS involved human participants (Beckers & Homberg, 1992).

We used information theory as a conceptual tool to introduce our framework. The conceptual dimensions of sensitivity, specificity, and invariance describe the relation between *s*, *n* and *r*. Functionality describes the use of *r* in downstream cognitive processes and the production of *b*. Together, these conceptual dimensions can be used to describe the typical empirical findings on the basis of which claims about representation are made in cognitive neuroscience.

# 3. Our framework applied across methodologies

We turn to a survey of common data analysis methods. We consider how analyses of correlation, decoding and encoding models, tests of statistical dependence, Representational Similarity Analyses, and findings about adaptation are used to provide evidence about representations and how this is systematized by our framework. We highlight formal connections between the information-theoretic formalism of our framework (Fig. 2) and these methods, to explain how they are related conceptually, but we do not work out in general under what conditions these methods would be mathematically equivalent. This discussion explains how our framework applies to a broad range of commonly applied methods but is only a starting point of an analysis of the full range of methods applied in cognitive neuroscience.

## 3.1. Information theory in neural data analysis

Information theory can be applied directly in neural data analysis. This may be motivated by not only thinking of the brain as an information processing system but also the utility of information theory for evaluating statistical relations between variables without assumptions about the structure of those relations (its model independence) (Borst & Theunissen, 1999; Ince et al., 2017; Nemenman et al., 2004; Ostwald & Bagshaw, 2011; Panzeri et al., 2008; Perkel & Bullock, 1968; Victor, 2006). For instance, information theory has been used to characterize the manner in which spike trains carry information about the stimulus (Bialek et al., 1993; Rieke et al., 1993), the effects of attention on population codes (Saproo & Serences, 2010; Serences et al., 2009) and object representations (Guggenmos et al., 2015), the relation between oscillations in EEG data and face expression perception (Schyns et al., 2011), the relation between EEG and fMRI data (Caballero-Gaudes et al., 2013), or to characterize the processing of rhythmic components of speech (Gross et al., 2013). Where entropies of variables and their mutual information are estimated, this affords the most straightforward interpretation in terms of our framework (Fig. 2).

Since estimates of entropy and mutual information generally require estimating full probability distributions (Box 1), rather than moments of the distribution like variance, they are particularly challenging for high-dimensional and continuous variables from limited sample sizes. Often, to obtain estimates of information-theoretic quantities, data is discretized into relatively few bins such that there is sufficient data for each bin. However, such binning introduces biases. Methodologies are being developed within neuroscience and in the larger context of statistics and machine learning with an eye towards benchmarking the reliability of estimators with respect to these challenges (Alemi et al., 2019; Álvarez Chaves et al., 2024; Kraskov et al., 2004; Piga et al., 2024) (see section 3.3 'Decoding models' for discussion as many such methods are based on them).

A particular challenge for continuous variables is the normalization of information-theoretic quantities. Take specificity, which quantifies how much of the total variability of $r$ carries information about $s$ (Fig. 2). In the discrete case, the total variability of a variable can be measured by its entropy. In the continuous case, no trivial measure of total variability is applicable to probability distributions in general (but see Nagel et al. (2024) for a discussion of general estimators for normalized mutual information).

The rate with which $r$ provides information to downstream processes may be used to normalize information-theoretic quantities (Supplementary Box 3). This rate is limited, for instance, by noise which may manifest in the temporal resolution with which downstream processes depend on $r$. Rieke et al. (1993) evaluated the mutual information between $s$ and $r$ ($I(s;r)$) normalized by the entropy of $r$ ($H(r)$) relative to a discretization of $r$ (with the size of the discrete bins of $r$ assumed to correspond to the lower limit of the temporal resolution of the neural code). While the normalization by $H(r)$ relies on a discretization of $r$ (potentially introducing biases), it does reflect not only stimulus-dependent variability but also endogenously driven processing and noise corrupting the signal. By contrast, Tafazoli et al. (2017) normalized $I(s;r)$ by the total information $r$ carries about the stimulus ($I(s,n;r)$). The normalization by $I(s,n;r)$ allows the use of a broader range of estimators for information-theoretic quantities that may avoid the biases of discretization and only captures the proportion of stimulus-dependent variability driven by $s$ or $n$ but not noise or other stimulus-independent drivers of variability.

## 3.2. Linear correlation

A very common quantitative measure of the relationship between two variables, whether discrete or continuous, is their Pearson correlation. High correlation between $s$ and $r$ means that $s$ can be predicted well from $r$ and that $r$ can be predicted well from $s$ (both sensitivity and specificity are high). As correlation is symmetric, it does not distinguish between sensitivity and specificity, which are asymmetric (when $r$ is highly sensitive to $s$, it need not also be highly specific to $s$; Fig. 3).

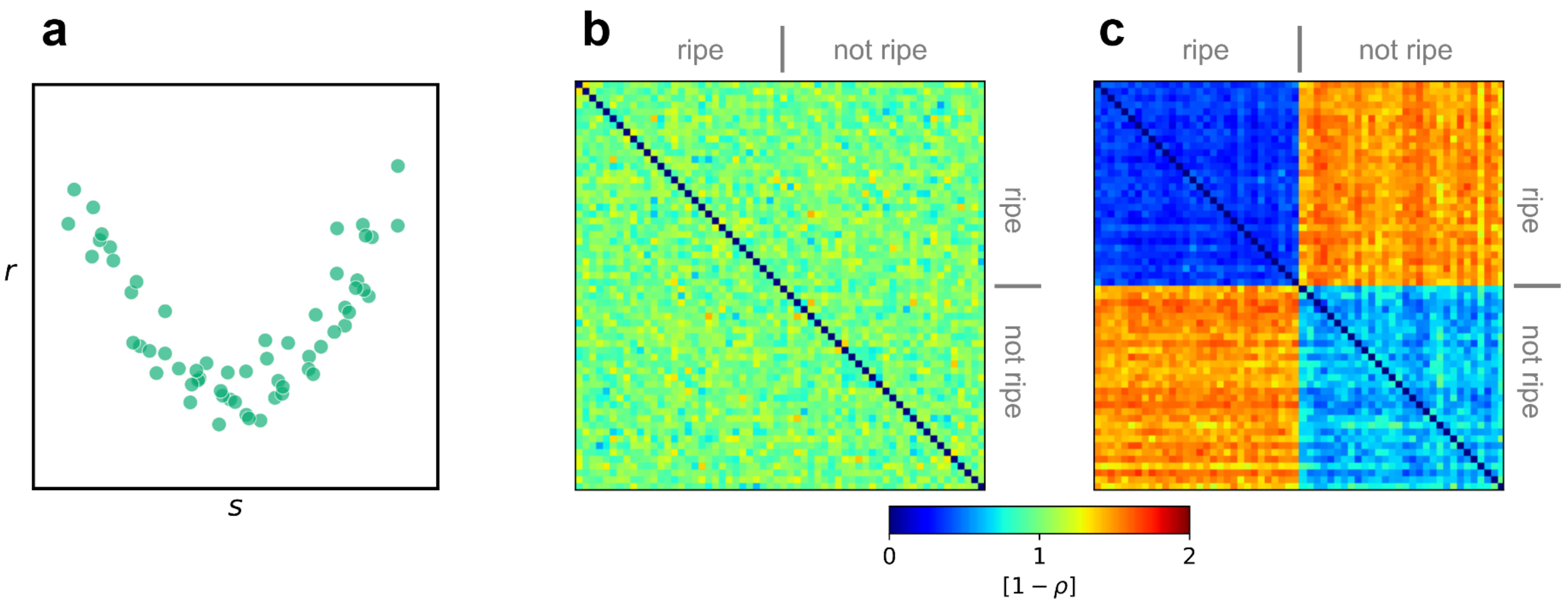

**Figure 4. How common measures are related to our framework. a** | A nonlinear relationship between *s* and *r* (*r* depends on *s* by a quadratic function). There is high mutual information between *s* and *r*, but *s* and *r* are not correlated. **b-c** | Interpretation of representational dissimilarity matrices (RDMs). In this hypothetical example, thirty ripe apples and thirty apples that are not ripe were presented to a participant. *r* was recorded from different brain areas (b and c) and correlations of *r* were computed for all pairs of stimuli (all pairings of apples, across all values of ripeness) to construct RDMs. There is no difference between dissimilarities for ripe and not ripe apples in the RDM for brain area b; *r* is not sensitive to the category distinction (**b**). By contrast, in the RDM for brain area c there is high dissimilarity across the category boundary of ripe versus not ripe; *r* is sensitive to this distinction (**c**). Yet there is little dissimilarity between exemplars within categories; hence *r* appears to be invariant to within category distinctions.

Lower degrees of correlation are more difficult to interpret, because mutual information may be high even though correlation is low when the dependence of *r* on *s* is non-linear (Reshef et al., 2011) (for example, U-shaped ; Fig. 4a). If the dependence between r and s is linear with added Gaussian noise, then the mutual information is monotonically related to the Pearson correlation coefficient (Cover & Thomas, 2006). In this scenario, low degrees of correlation also imply low mutual information and, therefore, low sensitivity and specificity.

Correlation may also be used to evaluate invariance, where a low degree of correlation of *r* with *n* indicates invariance to *n* (but again, *r* may in fact depend non-linearly on *n*). The correlation between *r* and *b* is an estimate of functionality — but only a crude one, since it poorly reflects causal dependence.

## 3.3. Decoding models

The most common measures of sensitivity are measures of the performance of a decoding model, that is, a model that infers *s* from *r*. Such models have been used to decode faces (Abbott et al., 1996) and object identity (Hung et al., 2005) from the monkey temporal cortex, to decode spatial location from the human hippocampus (Hassabis et al., 2009), auditory location from the monkey auditory cortex (Miller & Recanzone, 2009), and object category from early and late visual areas in humans (Cox & Savoy, 2003).

The accuracy of a decoding model is a good measure of sensitivity, because it has an easy to interpret lower and upper bound. When the decoding accuracy is at the chance level, this means that the given decoding model cannot extract any information about *s* from *r*, suggesting that the sensitivity is 0. The higher the decoding accuracy is above the chance level, the higher the sensitivity, with 100% decoding accuracy meaning one can perfectly infer *s* from *r* (the sensitivity of *r* to *s* is 1).

A general measure of sensitivity can be derived from the expected logarithmic loss of a decoding model estimated by the cross-validated loss on a separate test dataset, rather than the loss on training data. Consider a decoding model $p_\theta(s|r)$ with parameters θ in which $E[-\log p_\theta(s|r)]$ is its expected logarithmic loss under the true joint probability distribution $p(s,r)$. If the decoding model is fitted well and $p_\theta(s|r)$ is identical to $p(s|r)$, then $E[-\log p_\theta(s|r)] = H(s|r)$ (that is, the expected logarithmic loss of the decoding model is an estimate of the conditional entropy $H(s|r)$). An expected logarithmic loss of 0 means that the sensitivity of *r* to *s* is 1; *s* can perfectly be decoded from *r*. That is, the lower bound on the expected logarithmic loss of a decoding model corresponds to the upper bound on sensitivity. The lower bound on sensitivity corresponds to the chance level of a decoding model — when the sensitivity is 0, the expected logarithmic loss is equal to the loss of a model based on the prior probability distribution of *s* ($p(s)$) alone ($E[-\log p(s)]$), which estimates $H(s)$. Sensitivity can be estimated from the expected logarithmic loss of a decoding model as follows:

$$\text{Sensitivity: } \frac{I(s;r)}{H(s)} = \frac{H(s)-H(s|r)}{H(s)} = \frac{\mathrm{E}[-\log p(s)]-\mathrm{E}\left[-\log p_\theta(s|r)\right]}{\mathrm{E}[-\log p(s)]}$$

Any measure of a decoding model's performance that covaries with the expected logarithmic loss can be used as a measure of sensitivity in an analogous manner. Expected logarithmic loss and accuracy of a decoding model are useful particularly for classification problems (decoding of discrete variables). The mean squared error or absolute error are common measures in the case of continuous variables, where decoding models commonly only offer a point estimate for *s*, rather than an estimate of the full probability distribution $p(s|r)$.

By contrast, when $p_\theta(s|r)$ and $p(s|r)$ mismatch, then the expectation of the logarithmic loss of a decoding model merely offers an upper bound for the true entropy $H(s|r)$. That is, the estimate of sensitivity with a decoding model is merely a lower bound on the true sensitivity. A decoding model $p_\theta(s|r)$ with parameters θ may fail to closely approximate the true probability $p(s|r)$ for two main reasons. First, the available dataset (both training and test data) may mismatch the underlying distribution because it is too small or biased. Second, the decoding model may not be sufficiently complex to fit the true relationship between *s* and *r* (for example, a linear model will not pick up on non-linear relationships). A decoding model may also suffer from excessive complexity, especially with small datasets. Bayesian decoding includes explicit measures of model complexity (to avoid overfitting) and of how well a decoding model fits the data (Friston et al., 2008).

As the sensitivity estimated with a decoding model is a lower bound on the true sensitivity, by showing that a decoding model performs better than chance ($E[-\log p_\theta(s|r)] < E[-\log p(s)]$), one shows that the sensitivity of *r* to *s* must be larger than 0. However, comparative claims are more difficult to interpret. Even when the sensitivity of *r* to *s*, estimated with a decoding model, is larger than the sensitivity of *r* to *n*, this does not necessarily mean that *r* is truly more sensitive to *s* than to *n* because one might have underestimated the sensitivity of *r* to *n*.

Two perspectives on decoding models can be distinguished. First, the decoding model can be taken as a model of the statistical relation between variables and as a tool for quantifying sensitivity. Since more complex models — with sufficient precautions to avoid overfitting — tend to be more flexible in approximating statistical relations than simpler models, this perspective motivates the use of complex machine learning tools like support vector machines (Cortes & Vapnik, 1995) or more unconstrained deep neural networks (Livezey & Glaser, 2021). A second perspective interprets the decoding model not just as a model of a statistical relation but as a model of the process by which information about *s* is read from *r* within the brain. Under this perspective, the complexity of decoding models is often deliberately restricted, for instance to linear models, because it is assumed that linearly decodable features can be read by a single downstream neuron or layer (DiCarlo & Cox, 2007; Kriegeskorte & Diedrichsen, 2019). In this perspective, the decoding model is not simply a tool for estimating sensitivity but also a model of the functionality of *r* that aims to uncover the neural code that is being used in the brain (Supplementary Box 1).

## 3.4. Encoding models

An encoding model predicts *r* from *s* and *n* (for example, by modeling $p(r|s,n)$). Encoding models can be used to estimate specificity analogously to how decoding models are used to estimate sensitivity. In addition, by determining the role *n* has in an encoding model of *r*, invariance to *n* can be demonstrated (*r* is invariant to *n* given that *s* and *n* predict no more of *r*'s variability than *s* alone (Fig. 3c)). Examples of encoding models include models of the receptive fields of neurons in the cat visual cortex (Hubel & Wiesel, 1959, 1962), models of shape-tuning in populations of neurons in monkey V4 (Pasupathy & Connor, 2002), and population receptive field models with fMRI of the human visual cortex (Dumoulin & Wandell, 2008). General linear models in fMRI analysis (Friston et al., 1994, 1995) are also encoding models, as are generalized linear models (Paninski et al., 2007) and linear-nonlinear models (Aljadeff et al., 2016) in the analysis of electrophysiology data.

Analogous to decoding models, the expected logarithmic loss of an encoding model that predicts *r* from *s* ($E[-\log p_\theta(r|s)]$), can be used to estimate the specificity of *r* to *s*.

Specificity: $\frac{I(s;r)}{H(r)} = \frac{H(r)-H(r|s)}{H(r)} = \frac{\mathrm{E}[-\log p(r)]-\mathrm{E}\left[-\log p_\theta(r|s)\right]}{\mathrm{E}[-\log p(r)]}$

The lower bound on the expected logarithmic loss of the encoding model corresponds to the upper bound on specificity; when *r* can perfectly be predicted from *s* — E[-log $p_\theta$(*r*|*s*)] = 0 — then specificity is 1. When the expected logarithmic loss is above the baseline predictability of the neural population without knowledge of *s* (E[-log *p*(*r*)]) — E[-log $p_\theta$(*r*|*s*)] > E[-log *p*(*r*)] — then specificity is larger than 0. The baseline predictability is often assessed by shuffling the relationship between *r* and *s*, thereby setting the shuffled *r* and *s* to be independent and therefore removing any mutual information between *r* and *s* in the dataset.

As in the case of decoding models, encoding models can take many forms, from linear models with Gaussian noise to much more complex deep neural network based models (Kriegeskorte, 2015); and estimates of specificity established with them are merely lower bounds on the true specificity. Comparisons of specificity estimates, such as the claim that the specificity of *r* to *s* is larger than the specificity of *r* to *n*, are only as well supported as the assumption that one's encoding model captures all the dependencies between *r* and *s* and between *r* and *n*.

Encoding model performance can be measured with all the measures that are also available for decoding models. One that is common particularly for encoding models is the proportion of variance explained, which is similar to the information-theoretic quantification of specificity (Fig. 2, Box 1), except that the variability unexplained by *s* is measured with the variance of the residual of a model of *r* given *s* rather than the entropy:

Specificity: $\frac{I(s;r)}{H(r)} = \frac{H(r)-H(r|s)}{H(r)} = 1 - \frac{H(r|s)}{H(r)}$

Proportion of variance explained: $1 - \frac{var(r-m(r;s))}{var(r)}$,

where *m*(*r*;*s*) is a model that predicts *r* from *s*.

Variance is a computationally convenient measure of variability both for discrete and continuous variables that can be estimated from small datasets more robustly than entropy, but it has limitations. For instance, variance is not sensitive to the shape of probability distributions (such as whether they are skewed or multi-modal).

When *s* is continuous, an encoding model may also be used to estimate the sensitivity of *r* to *s* with the Fisher information (*J*(*s*)) (Brunel & Nadal, 1998; Fisher, 1925; Paradiso, 1988; Wei & Stocker, 2016), defined as $J(s) = \mathrm{E}[(d/ds \log p(r|s))^2]$. Intuitively, by considering the derivative of a (log-transformed) encoding model (d/d*s* log *p*(*r*|*s*)), *J*(*s*) quantifies locally around a given value of *s* how much of a difference small changes in *s* make for *r* and therefore how informative *r* is about *s*. *J*(*s*) is closely related to the signal-to-noise ratio (SNR) (Seung & Sompolinsky, 1993) and to the Signal Detection Theory measure of discriminability (*d'*), which quantifies how well two neighboring values of *s* can be discriminated based on *r* (Averbeck & Lee, 2006; Seung & Sompolinsky, 1993; Wei & Stocker, 2016). *J*(*s*) and related measures of discriminability (like SNR and *d'*) are evaluated commonly in analyses of shapes of tuning curves and noise correlations to study properties of the neural code for *s* (Averbeck & Lee, 2006; Bejjanki et al., 2011; Haefner & Bethge, 2010; Pouget et al., 1999; Rumyantsev et al., 2020; Seung & Sompolinsky, 1993; Yoon & Sompolinsky, 1998; Zhou et al., 2024), but also in analyses abstracting away from tuning curves (Kriegeskorte & Wei, 2021; Wei & Stocker, 2016). By the Cramér-Rao bound (Cramér, 1946; Rao, 1945), which states that the inverse variance of any unbiased decoding model of *s* from *r* cannot exceed $J(s)$, $J(s)$ can be used to establish an upper bound on decoding performance (sensitivity), based on an encoding model. Evidence for functionality is provided by demonstrating that measures of how well *s* can be discriminated given *r* (based on an encoding model) match measures of how well *s* is discriminated by *b*. It was shown, for instance, that the minimal differences in orientation humans could reliably discriminate matched the minimal difference in orientation that could be discriminated based on an encoding model of orientation in V1 (Paradiso, 1988).

## 3.5. Representational similarity analysis

Representational similarity analysis (RSA) seeks to build an understanding of low-dimensional geometric properties of high-dimensional neural population activities and their dependence on stimulus features (Kriegeskorte, Mur, & Bandettini, 2008). RSA is related to and builds upon analyses of representational geometry more generally (Cunningham & Yu, 2014; Shepard, 1980; Umakantha et al., 2021), including multidimensional scaling (Hasselmo et al., 1989; Young et al., 1995; Young & Yamane, 1992), and principal component analysis (Huth et al., 2016; Nicolelis et al., 1995).

In RSA, the geometry of neural representations is captured by computing a representational dissimilarity matrix (RDM). The RDM is constructed by computing dissimilarity in population responses to sets of stimuli. Given a set of *N* stimuli ($S_1$, …, $S_N$), an *N* x *N* dissimilarity matrix can be constructed, where each entry $d_{ij}$ is calculated by computing the dissimilarity in the neural population response to stimuli $S_i$ and $S_j$ (Kriegeskorte, Mur, & Bandettini, 2008) (Fig. 4b,c). The most used dissimilarity measure is the correlation distance, defined as 1 minus the correlation between the population responses to $S_i$ and $S_j$, computed across neurons or voxels. Alternative dissimilarity measures have been proposed, like classification accuracy of decoding models, Euclidean distance between values of *r* (Walther et al., 2016), or measures based on non-linear transformations of *r* (Lin & Kriegeskorte, 2024).

Typically, when computing the RDM, each stimulus is presented multiple times and the average neural population response to each stimulus is used to compute dissimilarity. This stimulus-conditioned averaging of the neural population response averages out the non-stimulus-dependent variability. As a result, RDMs tend to poorly reflect the specificity of *r* to the stimulus set, and to more closely reflect its sensitivity.

RDMs can be used to evaluate sensitivity and invariance. If *r* carries no information about the stimulus, the dissimilarity of neural activity patterns will be the same, independently of the stimulus they were measured in response to. The resultant RDM would be unstructured and uniformly dominated by noise (Fig. 4b). Since such an unstructured RDM suggests zero mutual information between *r* and *s* it also suggests that sensitivity is 0. Conversely, if a neural population carries information about the stimuli, we expect the RDM to display a structure reflecting the dissimilarity of stimuli. Often, stimuli are drawn from categories with multiple example stimuli present from each category. Dissimilarity of stimuli across category boundaries (such as houses versus faces) indicates sensitivity to the category distinction. Low dissimilarity of stimuli within categories indicates invariance to within-category distinctions (such as variations of the orientation of a face; Fig. 4c). Similarity of RDMs computed from *r* and RDMs computed from *b* can also provide evidence for the functionality of the neural response (Op De Beeck et al., 2001).

Perhaps the most powerful application of RSA lies in comparisons across systems, such as brains of different individuals, different brain areas, brains of different species (Kriegeskorte, Mur, Ruff, et al., 2008), or even brains and artificial neural networks (Li et al., 2019; Yamins et al., 2013). Yet such comparisons are not primarily concerned with what evidence establishes that a feature is represented — the focus of this Perspective — but with studying how structures of representations compare across systems.

## 3.6. Tests of statistical dependence

Some analyses do not estimate the amount of information between variables and, therefore, do not offer quantitative estimates of the conceptual dimensions of representation. Rather, they test hypotheses about relations of statistical dependence, which suggest that respective dimensions are either satisfied or fail. For instance, initial work demonstrating that the fusiform face area (FFA) is sensitive to faces showed that fMRI measurements of activity in FFA were significantly higher in conditions in which an image of a face was shown to a participant compared to another object like a house (Kanwisher et al., 1997). Of note, this established that FFA carries information about faces without quantifying how much information.

Establishing mutual information between *r* and *s* also establishes that *r* has larger than 0 sensitivity and specificity to *s*. That FFA responded to faces irrespective of whether the face is viewed from the front or from the side suggests invariance to viewing angle. That FFA did not respond to other object categories like houses suggests specificity to faces. Yet such interpretations in terms of invariance and specificity have to be seen with caution as they rest on a null-finding.

Many tests for relations of statistical dependence merely look for a dependence between *s* or *b* and the mean neural response of an isolated element (for example, an individual neuron or a voxel in a fMRI analysis) in a so-called univariate analysis. Such tests cannot pick up on the information carried by activity patterns distributed across neural populations. To pick up on such patterns, a multivariate pattern analysis in terms of correlation or more complex encoding or decoding models is needed (Haxby et al., 2001, 2014).

## 3.7. Adaptation

The conceptual dimensions of representation may also be evaluated using the phenomenon of adaptation. Upon repeated presentation of an unchanging feature, the response magnitude of neurons (Baylis & Rolls, 1987; Sobotka & Ringo, 1994) or brain regions (Grill-Spector et al., 1999; Weigelt et al., 2008) that are sensitive to the feature typically decreases. This adaptation is disrupted when the feature changes. Hence, the established computational or physiological mechanism of adaptation can be used to probe the characteristics of a representation. When adaptation is disrupted by changes in $s$, $r$ is sensitive to $s$. When adaptation persists across changes of $n$, $r$ is invariant to $n$; as far as $r$ is concerned, stimuli that differ in $n$ are still equivalent.

To illustrate, fMRI activity in the lateral occipital complex (LOC) adapts to presentations of the same object (Grill-Spector et al., 1999). Changes in illumination or viewpoint more strongly disrupted this adaptation than changes in position or size of the object. That is, LOC shows some invariance to the position or size of an object, but it is less invariant (more sensitive) to its illumination and viewpoint.

In summary, information theory may be applied in neural data analysis, which affords the most straightforward interpretation in terms of our framework. Yet we have also explained how other common analysis methods like linear correlation, decoding and encoding models, tests of statistical dependence, Representational Similarity Analysis, and findings about adaptation may be used to evaluate the conceptual dimensions of representation of sensitivity, specificity, invariance, and functionality.

# 4. Canonical examples

In our view, neuroscientific research about representations has always been guided implicitly by the framework we have laid out. Our framework is supposed to capture, in a systematic and unified way, what researchers have been doing all along. We look at a few examples of lines of research. These examples were chosen arbitrarily based on having received high attention from researchers and having been involved in developing and shaping the tools and analyses with which representations in the brain are being studied. We do not offer exhaustive reviews but focus on illustrating how tests of the conceptual dimensions of representation combine to form comprehensive bodies of evidence. A single study may not address all conceptual dimensions, but as a line of research matures, evidence across all of them tends to accumulate.

## 4.1. Orientation of visual elements

Research about the visual representation of orientation began with characterizing the response profiles of cells in the primary visual cortex (V1) (Hubel & Wiesel, 1959, 1962, 1968). Researchers used encoding models in terms of tuning curves — which estimate the level of activity (and variance) as a function of $s$ like orientation — to evaluate specificity to orientation (Sompolinsky & Shapley, 1997; Victor et al., 1994). While the V1 population of neurons is sensitive not only to orientation but also to other features like contrast, spatial frequency, and spatial phase of oriented gratings, some sub-populations like complex cells displayed invariance at least to spatial phase while maintaining sensitivity to orientation (De Valois et al., 1982). In both simple and complex cells, the preferred orientation and width (but not the height) of tuning curves was invariant to contrast (Anderson et al., 2000; Hansel & van Vreeswijk, 2002) and temporal frequency (Moore et al., 2005). Yet, at least the contrast invariance of the width of tuning, may depend on adaptation to contrast (Nowak & Barone, 2009).

On the level of population responses, sensitivity of V1 to orientation has been characterized further with decoding models (both linear and nonlinear) (Berens et al., 2012; Chen et al., 2006, 2008; Graf et al., 2011). Berens et al. (2012) showed that, even though individual neurons are not invariant to contrast, on the population level, orientation could be decoded invariant to contrast. Chen et al. (2006) evaluated conditional specificity by showing that even for very low contrast levels, the signal-to-noise ratio for responses to orientation was large.

That V1 response patterns are functional with respect to orientation has been tested by predicting *b* to oriented stimuli from V1 responses. The perceived orientation reported by monkeys with saccadic eye movements can be predicted above chance from single V1 neurons, even in trials where the stimulus contained no oriented signal, suggesting that the orientation percept is caused by random variations of V1 responses (Nienborg & Cumming, 2014). On the level of V1 population responses, classification of orientations by monkeys can be predicted, using a deep neural network decoder, even when the same stimulus is presented multiple times across trials (Walker et al., 2020), again, suggesting that behavioral responses are caused by random fluctuations in the V1 response across trials.

## 4.2. Numerosity

The numerosity of a collection of things is the number or cardinality of how many things there are in a collection.

On the one hand, initial observations of deficits in the processing of numerosities owing to lesions of the human parietal cortex suggested that the parietal cortex played a causal role in the processing of numerosities. That is, they suggested that there is a parietal representation of numerosity that is functional (Dehaene & Cohen, 1995). Intervention studies have subsequently confirmed this finding of functionality. The repetitive stimulation of parietal areas in humans with TMS impaired the processing of numerosity (Dormal et al., 2008) whereas pharmacological inactivation of numerosity sensitive neurons in the monkey's parietal cortex stopped behavioral responses to numerosity (Sawamura et al., 2010). In addition, early fMRI work with humans showed that parietal activity statistically depended on numerical distance in a task of comparing numerosities (Pinel et al., 1999) and neurophysiology with monkeys confirmed that the firing rate of parietal neurons statistically depended on numerosity (Nieder & Miller, 2004). The statistical dependence on numerosity implies that parietal neural responses carry information about and are sensitive to numerosity, while a lack of dependence on other factors suggests specificity to numerosity.

On the other hand, invariance has been an ongoing source of skepticism about whether numerosity is really represented in the parietal cortex (rather than a generic quantity that is a mixture of multiple quantitative features). We see here that the conceptual dimensions of representation can not only be used in support of but also for criticism of claims about representations. Invariance of parietal cortex representations has been demonstrated to features including low level visual features (Nieder et al., 2002; Nieder & Miller, 2004) symbolic and non-symbolic (Pinel et al., 1999) and visual versus auditory presentations of numerosities (Nieder, 2012) suggesting that it is numerosity specifically that is being represented. However, controversy remains with evidence suggesting a lack of invariance especially to spatial features like size. In one study (Harvey et al., 2013), encoding models with tuning to preferred numerosities were fit to fMRI data and the preferred numerosity depended on item size. Another study (Pinel et al., 2004) has found that neural activity in overlapping areas of the parietal cortex depended on both numerosity and size of Arabic numerals used to present the numerosity. This lack of invariance suggests that, perhaps, not numerosity specifically, but a mixed feature that combines spatial size and numerical size is represented. These conflicting findings may be reconciled if there is a distributed population code for numerosity and related quantities (Tudusciuc & Nieder, 2007) that supports invariant readouts of numerosity by downstream neural processes only in task specific settings. When numerosity is not task relevant, a mixed quantitative feature may be represented: only when it is task relevant, may neurons be tuned to represent numerosity specifically (Castaldi et al., 2019).

## 4.3. Spatial Location

Place cells are cells in the hippocampus that respond with increases in spiking activity when the animal is in a particular location in the environment, discovered in 1971 by O'Keefe and Dostrovsky (1971). The sensitivity of place cells to spatial location is conditional on the presence of visual cues, but is invariant to any particular visual cue like oriented gratings moving in the rat's visual field (O'Keefe & Dostrovsky, 1971), as well as further environmental cues, including olfactory and auditory cues (Battaglia et al., 2004; Jeffery & O'Keefe, 1999; Sharp et al., 1990). As place cells do not fire in response to modulations of such environmental cues, they are also specific to spatial location.

Across the whole hippocampus population of cells, the animal's position for all investigated locations can be inferred (Wilson & McNaughton, 1993), suggesting that there are not just place cells sensitive to a few specific locations, but that the population response is sensitive to location across the animal's environment. Some place cells had specificity to spatial location in a relative reference frame defined by reward location or landmarks that was invariant to absolute location in the experimental environment, whereas others showed specificity to absolute location and invariance to changes in reward location and landmarks (Gothard et al., 1996). Sensitivity to space across the environment in an absolute reference frame has been taken to suggest that the animal represents a map of its environment (O'Keefe & Nadel, 1978).

Stimulation of place cells sensitive to locations previously associated with reward or starting position induced behavior appropriate for those locations, even when the rat was not located there (Robinson et al., 2020). Rewarding stimulation of the median forebrain bundle correlated with spontaneous place cell activity during sleep to induce associations between specific locations and reward also induced goal directed behavior by the rat toward those locations when placed in the experimental environment again after awakening (De Lavilléon et al., 2015). Both interventions suggest that place cells play a causal role in producing location specific behavior and, therefore, are functional as representations of spatial location.

In all three examples, the representation of orientation in V1, the representation of numerosity in the parietal cortex, and the representation of spatial location in the hippocampus, all four conceptual dimensions of representation have been addressed in the literature. Each exemplify mature lines of research. Still controversies may remain. Individual neurons in V1 lack invariance to features like contrast. The population response affords invariant readouts of orientation, but the mechanisms by which downstream processes read information about orientation from V1 require further investigation. The invariance of parietal representations of numerosity to spatial size remains controversial as well. Yet while there may be disagreement, for instance, about the extent to which invariance is satisfied in some of these cases, there is no disagreement over whether invariance is relevant for whether a feature is represented. We believe these examples show that the evidence that is relevant for claims about representation in neuroscience is systematized well by our framework.

# 5. Extensions of our framework of conceptual dimensions of representation

For ease of exposition, simplifications have been made in our discussion of the framework. We want to acknowledge some of the most notable limitations and suggest extensions.

First, there is more to say about *r*. Implicitly, we have treated *r* as a maximally detailed measure of activity across a population of neurons. However, some work is concerned more specifically with modeling what aspect of a population response is relevant for representing *s*. A classic question concerns whether the timing of spikes or merely the firing rate matters (Bialek et al., 1991; London et al., 2010; Rieke et al., 1999). A related question is how a single population of neurons displays mixed selectivity (where different aspects of *r* may represent different features $s_1$, $s_2$, ...) (Fusi et al., 2016; Rigotti et al., 2013). Our framework can also be applied to models of the relevant aspect of *r* (Supplementary Box 3 explains its application for this purpose).

Second, there is more to say about values of *s* and how they relate to values of *r*. One may not only want to assign *s* as the content of *r* but also model what value of *s* is represented by a particular activity pattern of *r*. That is, one may want to be able to read the neural code (Supplementary Box 1 explains the application of our framework to models of the neural code).

Besides the relaxation of these two simplifications, there are a few more caveats we want to consider.

Neuroscientific work on representations is often an extension of behavioral psychological research. Behavioral work identifies which features are represented (it identifies *s*), and neuroscientific work localizes these representations in the brain (it identifies *r*). Where neural data is particularly difficult to collect, for instance in social cognition or developmental psychology, which focuses on humans, great apes, or children, most research on representations is based on behavioral data alone. Without estimates of neural variability, specificity cannot be applied. Yet models based on behavioral data do aim to identify representations that are sensitive to *s*, they reveal interferences between representations (failures of invariance), and, inherently, they also show that a representation is functional by being used in the production of *b*.

Further, the information-theoretic quantifications of the conceptual dimensions of representation (Fig. 2) reflect not only the dependencies between variables, but also the distributions of the variables in isolation ($p(s)$, $p(n)$, and $p(r)$). For instance, naturally distributed stimuli lead to higher estimates of the specificity of spike trains coding for auditory stimuli than unnaturally distributed stimuli (Rieke et al., 1995). This suggests that the neural code is optimized for the natural statistics of features. However, it is not generally the case that the conceptual dimensions of representation are more informative when evaluated under natural statistics. Often stimuli are deliberately drawn from unnatural distributions which allows teasing apart features that tend to covary in natural environments. For example, in the study of representations of numerosity, stimulus distributions are crafted in which features such as numerosity, item size, and item spacing vary independently as much as possible (DeWind et al., 2015).

Finally, we want to acknowledge that in this Perspective the discussion applies our framework only to representations that are about features of the participant's environment. Elsewhere we have discussed how our framework applies to representations of uncertainty (Walker et al., 2023). Uncertainty is special because of its observer relativity — it is a feature of a belief or representation an observer has about the world. To cover neuroscientific work in its full breadth one would also have to consider, for instance, representations of motor commands, goals, values, or imaginings.

# 6. Concluding remarks

Despite conceptual and terminological ambiguity, we believe there is implicit agreement in neuroscience on what is characteristic of representation. Our formal framework aims to make that agreement explicit. It does so by disambiguating and formalizing four kinds of relations between features, neural responses, and downstream effects of a neural response characteristic of representation that, previously, have not been teased apart systematically. We propose that researchers use the framework developed in this paper to describe their findings about the representations they investigate — and to explain how their data analysis methods yield estimates of the respective conceptual dimensions of representation. Using common terminology in this way would facilitate communication across research groups, would make it easier to see how different research approaches combine, and would afford easier meta-analyses. Few studies provide evidence for all the conceptual dimensions of representation; strong evidence for representation emerges only out of a combination of studies.

Following the framework we present in this Perspective makes more salient which evidence is missing from a line of research; it also allows researchers to determine when strong evidence for a representation has emerged in a field.

As measurement techniques develop, datasets get larger, and analysis methods get more complex, information-theoretic quantities can be estimated more robustly. Deep neural networks are particularly useful tools that need not be interpreted as models of the brain. In some cases, they may simply be tools for estimating conditional entropies, affording quantitative estimates of the conceptual dimensions of representation for different features allowing for richer comparisons across studies (see sections 3.3 'decoding models' and 3.4 'encoding models' for discussion).

How does our framework relate to philosophical discussions of representation? It is intended to systematize how representations are studied in neuroscience and is not designed to address philosophical questions about what a representation is. But, of course, how representations are studied in neuroscience is highly relevant for philosophical accounts of what a representation is. It will therefore be relevant to consider how philosophical accounts of representation interface with our framework.

Finally, interest has recently arisen in explaining artificial neural networks in representational terms. On the one hand, neuroscience benefits from developments in artificial intelligence research because that research provides new analysis and modeling tools for neuroscience. On the other hand, analysis tools for understanding artificial neural networks are being compared to neuroscientific methodology (Ivanova et al., 2021; Lindsay & Bau, 2023). Our framework is a natural starting point for such a discussion because it encapsulates the methodology that has developed over more than hundred years in neuroscience and the information-theoretic formalism lends itself naturally to an application to artificial neural networks.

# 7. Funding statement

FM is supported by an ERC grant 947105-NEURAL-PROB.

# 8. Author contributions

All authors contributed substantially to the development of our framework and to discussion of the content. SP wrote the paper with contributions from EW, DB, and JL. SP, RD, FM, and WM revised and edited the manuscript.

# 9. Supplementary material

**Supplementary Box 1: Reading the neural code**

A model of the neural code is a mapping $g$ from $r$ to estimates of $s$ ($\hat{s}$). $g$ may for instance map $r$ to a point estimate ($g(r) = \hat{s}$) (Bialek et al., 1991), or, in a probabilistic population code, $g$ may map $r$ to a probability distribution over possible values of $\hat{s}$ ($g(r) = p(\hat{s}|r)$) (Jazayeri & Movshon, 2006; Ma et al., 2006). A model of the mapping $g$ allows us to not only say what feature is being represented by $r$, but we can say for every particular value of $r$ (an activity pattern across a population of neurons) what particular value the feature is represented to take.

Formally, $g$ is a decoding model. However, not any decoding model is a model of the neural code. As is discussed in the section on ‘Decoding models’, many decoding models are mere analysis tools for estimating how much information $r$ carries about $s$. There is no commitment that such decoding models are models of any structure within the brain. $g$, on the other hand, captures how information about $s$ is read from $r$ by downstream operations within the brain.

If we distinguish between $r$ and aspects of $r$ relevant for representing $s$ (see Supplementary Box 3), we may decompose $g$ into, first, a mapping $f$ from $r$ to its relevant aspect, and secondly, a mapping $g'$ that assigns a feature estimate $\hat{s}$ or $p(\hat{s}|r)$ to $f(r)$ (that is, $g(r) = g'(f(r))$). Here $g'$ may be understood as an interpretation function, or a content assignment that assigns a content — an estimate of a feature — to $f(r)$.

Analogous to $f$, the search for $g$ is constrained by the conceptual dimensions of representation. We are looking for that $g$, such that $g(r)$ is sensitive to $s$, specific to $s$, invariant to $n$, and functional.

Since $g$ captures how information about $s$ is read from $r$ within the brain, functionality plays a special role (Panzeri et al., 2017; Perkel & Bullock, 1968; Pica et al., 2017). We for instance expect that $g$ allows us to predict, on a trial-by-trial basis, what the exact behavioral report of a feature will be. If, in a particular trial, the orientation of a stimulus is represented by $r$ to be thirty degrees right of vertical ($g(r_i)$ = 30°), then we expect the participant to also report it to be thirty degrees right of vertical. On the other hand, when the participant makes an error and reports the orientation to be 25 degrees when, actually, it is 30 degrees right of vertical, we expect to be able to explain the error in this particular trial either in terms of a misrepresentation of the orientation ($g(r_i)$ = 25°), or in terms of some interference by a process downstream of the representation.

Short of building an exact model of *g*, we may constrain the class of models that are plausible. Constraints on biological plausibility derive from anatomical and other biological considerations. Further constraints are derived from the successes and failures of encoding or decoding. For instance, when models are successful only with a certain kind of structure — for example, multivariate linear models — then the neural code also must have that kind of structure.

One salient family of constraints are constraints on the complexity of the neural code. It has for instance been suggested that representations — or at least explicit representations — must be linearly decodable (DiCarlo & Cox, 2007; Kriegeskorte & Diedrichsen, 2019). Such a constraint may be motivated by the observation that a linear code is easily readable by an individual neuron.

We are skeptical of a priori constraints on the complexity of the neural code. Simple neural codes like linear ones are useful because they are easily readable. So, we may find that some features, especially features that are evolutionarily important to an organism, may be represented in terms of linear codes. However, the brain is a deep neural network and can approximate codes of arbitrary complexity. Populations of neurons can perform computations over features that are not linearly decodable; already individual neurons compute nonlinear functions of their inputs. Constraints on the complexity of neural codes have to be discovered empirically.

**Supplementary Box 2: Mutual information between *r* and interactions of *s* and *n***

The mutual information between three or more variables can take counter-intuitive values when there are complex interactions. *r* may for instance carry more information about *s*, conditional on *n*, than about *s* alone. When applying the framework to variables with complex interactions, the conditional variants of sensitivity and specificity, as well as invariance, are of particular relevance.

A classic example is the case of XOR (Griffith & Koch, 2014). Let *s*, *n*, and *r* be binary variables. Let *r* be $r$ = XOR($s$;$n$). That is, *r* takes the value 1 if and only if $s = 1$ and $n = 0$, or if $s = 0$ and $n = 1$.

In this case, intuitively, *r* represents something about *s* and *n*, and yet the sensitivity of *r* to *s* and the sensitivity of *r* to *n* are both 0 ($I(s;r)/H(s) = I(n;r)/H(n) = 0$).

Conditional on *n*, however, we see that *r*'s sensitivity to *s* is 1 and, conditional on *s*, *r*'s sensitivity to *n* is 1 ($I(s;r|n)/H(s|n) = I(n;r|s)/H(n|s) = 1$). So, *r* represents something about both *s* and *n*. And yet *r* not only lacks sensitivity to *s* and *n*, it also lacks invariance to *n* ($1 - I(n;r|s)/H(r|s) = 0$) and invariance to *s* ($1 - I(s;r|n)/H(r|n) = 0$). So, it cannot be a representation of either *s* or *n*.

More clarity is found once we include XOR(*s*;*n*) in the analysis. While *r* lacks sensitivity to *s* and *n* in isolation, it is sensitive to XOR(*s*;*n*) ($I(\mathrm{XOR}(s;n);r)/H(\mathrm{XOR}(s;n)) = 1$) and it is specific to XOR(*s*;*n*) ($I(\mathrm{XOR}(s;n);r)/H(r) = 1$). Relative to its sensitivity to XOR(*s*;*n*), *r* is also invariant to both *s* ($1 - I(s;r|\mathrm{XOR}(s;n))/H(r|\mathrm{XOR}(s;n)) = 1$) and *n* ($1 - I(n;r|\mathrm{XOR}(s;n))/H(r|\mathrm{XOR}(s;n)) = 1$).

So, ultimately, the conceptual dimensions of representation reveal the intuitively plausible result that *r* is a representation of XOR(*s*;*n*).

Whenever the conditional variants of sensitivity or specificity are greater than the unconditional variants, this suggests that *r* is sensitive to a combination of *s* and *n*. Extensions to classical information theory like the partial information decomposition (PID) framework have been developed to quantify the information between more than two variables and, in particular, the information carried only by the combination of multiple variables (Williams & Beer, 2010) and have been applied to analyze the flow of information between multiple variables in neuroscience (Celotto et al., 2023; Lemke et al., 2024; Pica et al., 2017; Varley et al., 2023).

**Supplementary Box 3: Reducing *r* to its relevant aspect**

We can distinguish between analyses that focus on the level of neural population responses and take *r* to be a maximally detailed measure of this response (constrained only by limitations of measurement techniques) and analyses that identify some more specific aspect of the neural response. Analyses of the second kind may for instance test hypotheses about whether only firing rates or also the timing of spikes carry information about features (Bialek et al., 1991; London et al., 2010; Rieke et al., 1999). Here we discuss how our framework applies to models of the relevant aspect of a neural response.

Let *r* be a maximally detailed measure of the response of a population of neurons (such as spike trains, or even more detailed measures like the dynamics of membrane potentials). Let *f* be a mapping that reduces *r* to its relevant aspect. As an example, *f* may map the population response to a vector of firing rates per neuron, disregarding the timing of spikes. When considering the response of large populations of neurons, often, *f* is a dimensionality reduction or a mapping from $r$ to a subspace of $r$, that gives a reduced set of dimensions such that only variations of $r$ along those dimensions are hypothesized to carry information about $s$.

In practice, the search for $f$ often co-evolves with models of $s$. Consider the search for the neural code for faces; Chang and Tsao (2017) find that neurons in different parts of the macaque brain are sensitive and specific to small numbers of features either of the appearance or the shape of faces. Their study advances new hypotheses about $f$ and $s$ in parallel.

With the slight modification of substituting $f(r)$ for $r$, the conceptual dimensions of representation also guide the search for $f$; that is, we try to find some $f$, such that $f(r)$ is *sensitive* to $s$, *specific* to $s$, *invariant* to $n$, and *functiona*l. As an example, for specificity we get the following information-theoretic measure by substituting $f(r)$ for $r$:

$$\frac{I(s;f(r))}{H(f(r))}$$

A central advantage we get out of this substitution is an increase in specificity: the specificity of $f(r)$ to $s$ will be greater than the specificity of $r$ to $s$. This is because, usually, there will be much more variability in $r$ than in $f(r)$, that is $H(r) \geq H(f(r))$, and specificity is normalized by the entropy of $r$ or $f(r)$ respectively. As a result, $s$ explains a larger proportion of the variability of $f(r)$ than of $r$; we may even hope that most of the variability of $f(r)$ can be explained by $s$. The specificity of $r$ to $s$, on the other hand, cannot usually be expected to be too close to 1.

For similar reasons, $f(r)$ may be more invariant than $r$. Say $r$ displays mixed selectivity and is sensitive to the two features $s_1$ and $s_2$. If $r$ indeed represents both features — rather than a single feature that is a mixture of $s_1$ and $s_2$ — there have to be two aspects of $r$ ($f_1$ and $f_2$) such that $f_1(r)$ represents $s_1$ and $f_2(r)$ represents $s_2$. These two aspects may for instance be the overall level of activity in a population of neurons vs. the activity relative to the tuning of individual neurons. $f_1(r)$ may represent $s_1$ with perfect invariance to $s_2$, because $f_1$ abstracts away from $s_2$-related variability of $r$, whereas $r$ lacks this invariance. Such invariance is found for instance in the representation of orientation in V1. The overall level of activity depends on contrast, yet the level of activity relative to the tuning of neurons carries information about orientation independently of contrast (Anderson et al., 2000; Berens et al., 2012; Hansel & van Vreeswijk, 2002).

Finally, to satisfy functionality, $f(r)$ cannot merely be a simplifying description of $r$ that accounts for all or most of the mutual information between $s$ and $r$. To be functional, $f$ has to capture the computationally relevant variations of $r$ that are used downstream in the brain, for instance, in the production of behavior.

This dependence on functionality is particularly relevant in applications of our framework to continuous variables. For continuous variables, there is no trivial measure of total variability. However, since $f(r)$ is constrained by functionality, $H(f(r))$ is upper-bounded by the rate with which information about $s$ is transferred to downstream operations — that is, $I(f(r);x)$ where $x$ is the downstream operation receiving information from $f(r)$. This upper bound may be used as the normalizing factor instead of $H(r)$. In practice, this transfer rate can rarely be estimated, such that proxies may be used. As is discussed in the section on 'Information theory in neural data analysis', Rieke et al. (1993) normalize by $H(f(r))$ where $f$ is a discretization of $r$ and bin sizes correspond to the temporal precision of the neural code. Rieke et al. do not determine the temporal precision of the neural code by actually looking at the temporal precision with which downstream operations can read information from $r$. Instead, they chose bin sizes such that smaller binnings would not substantially increase $I(s;f(r))$. It may be assumed that temporal variations of $r$ too small to affect what information about $s$ is being carried are below the threshold of the temporal resolution of the neural code.